\documentclass[a4paper,twocolumn,english,aps,prl,superscriptaddress]{revtex4}
\usepackage[T1]{fontenc}
\usepackage[latin1]{inputenc}
\usepackage{float}
\usepackage{graphicx,color}

\usepackage{amsmath}

\usepackage{color}
\makeatletter

\usepackage{babel}
\makeatother
\begin{document}
\preprint{This line only printed with preprint option}

\title{Optimal pumping of orbital entanglement with single particle emitters}

\author{Y. Sherkunov}
\affiliation{Physics Department,
Lancaster University,
Lancaster,
LA1 4YB, UK}
\affiliation{Department of Physics, University of Warwick, Coventry, CV4 7AL,
UK}

\author{N. d'Ambrumenil}
\affiliation{Department of Physics, University of Warwick, Coventry, CV4 7AL,
UK}

\author{P. Samuelsson}\affiliation{Division of Mathematical Physics, Lund University,
Box 118, S-221 00 Lund, Sweden}
\author{M. B\"uttiker} 
\affiliation{D\'epartement de
Physique Th\'eorique, Universit\'e de Gen\`eve, CH-1211 Gen\`eve 4,
Switzerland} \date{\today}

\begin{abstract}
  We propose a method for the optimal time-controlled generation of
  entangled itinerant particles, using on-demand sources in a
  conductor in the quantum Hall regime. This entanglement pump is
  realized by applying periodic, tailored voltage pulses to pairs of
  quantum dots or quantum point contacts. We show that the pump can
  produce orbital Bell pairs of both electrons and holes at the optimal
  rate of half a pair per pumping cycle. The entanglement can be
  detected by a violation of a Bell inequality formulated in terms of
  low-frequency current cross correlations.
\end{abstract}
\maketitle

Entanglement of itinerant electrons in mesoscopic and nanoscale
systems continues to attract great attention. Despite intense
effort,  a clear demonstration of the
generation, spatial separation and detection of entangled pairs of
electrons is still missing. Of particular interest would be an on-demand
source for entanglement
\cite{Samuelsson05,BeenakkerTitov05,Das06,Sherkunov09}, while a key
component in quantum information processing is the time-controlled
production of entangled flying quantum bits. 
A scheme for the dynamical
generation, or pumping, of orbitally entangled electron-hole pairs was
proposed by two of us in
\cite{Samuelsson05}.
However, as the proposed pump operated in the weak amplitude regime, it
produced on average much less than one Bell pair per cycle.
Subsequently 
it was shown 
that, for strong amplitude pumping of non-interacting particles, the
optimal production rate is half a Bell pair per cycle \cite{BeenakkerTitov05}. 
Optimal electron-hole pair
entanglement pumps have also been proposed, but without any scheme for entanglement
detection \cite{BeenakkerTitov05,Sherkunov09}.

An important step towards entanglement
pumping was taken experimentally by F\`eve {\it et al} \cite{Feve}, who realized a
single-particle on-demand source
in a conductor in the quantum Hall regime
\cite{Keeling06,VanNazBel07,Mosk08,Keeling08,Zhang09,Battista11}. 
Under ideal conditions, the
source produces exactly one electron and one hole per cycle. A scheme
for the production of pairs of orbitally entangled particles 
in different time bins based on
two on-demand sources in a double electronic Mach-Zehnder
interferometer was proposed in \cite{Olkhovskaya, Splettstoesser09}. 
The origin of the entanglement was
two-particle interference \cite{Sam04,Ned07,Sam09}, manifested by a
non-local two-particle Aharonov-Bohm effect. The
maximum production was 1/4 Bell pair per cycle, {\it i.e.\/} half the optimal rate.

\begin{figure}[t]
\centering
 \includegraphics[width=0.4\textwidth]{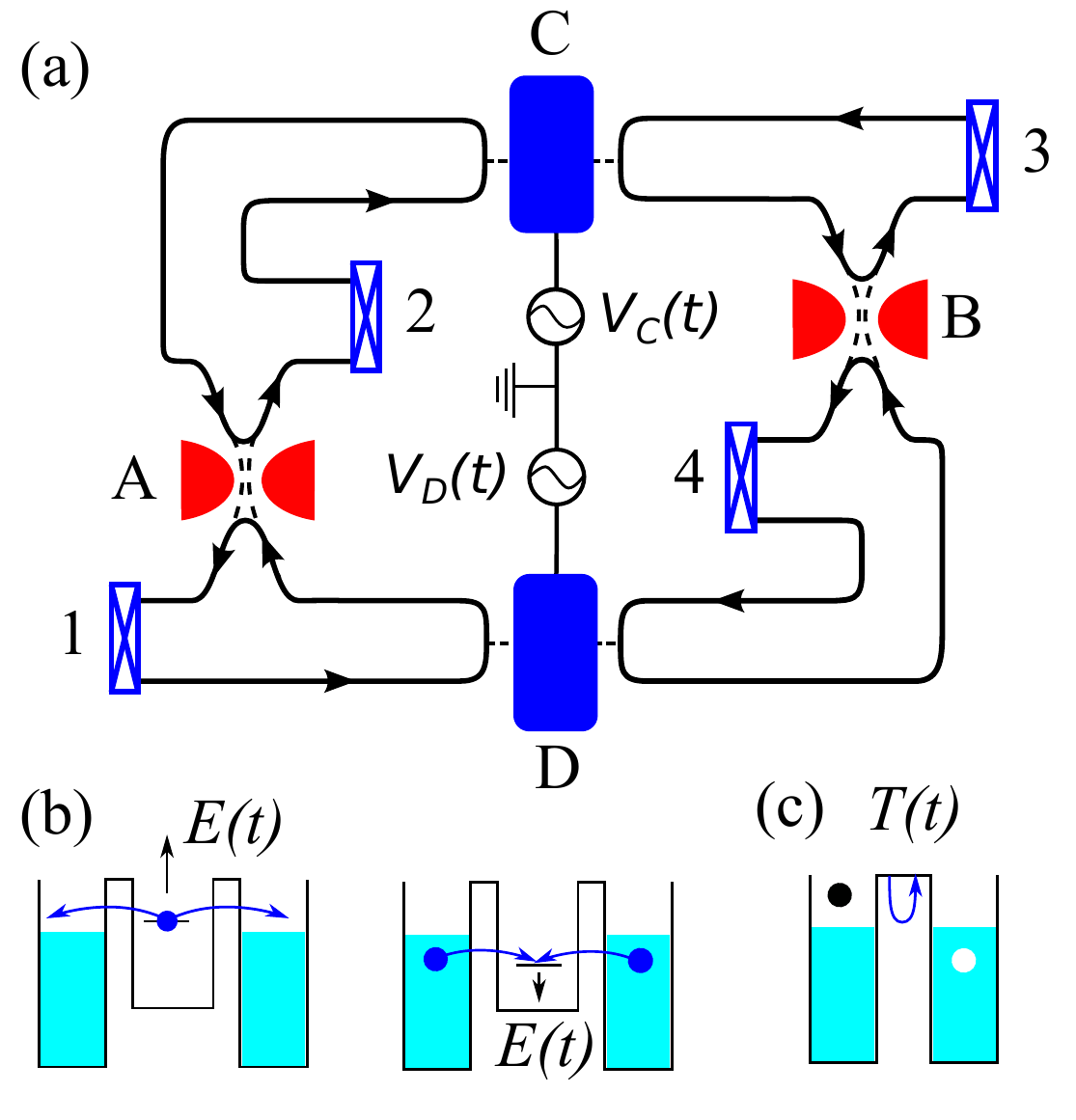}
 \caption{\label{fig:fig1} a) Optimal orbital entanglement pump with
   two single particle emitters, C and D, connected to conductors in
   the quantum Hall regime. The emitted electrons/holes propagate
   along edge states to controllable electronic beam splitters at A
   and B and are detected at terminals 1-4. The sources C and D can be
   either b) quantum dots or c) quantum point contacts. Tailored voltage 
   profiles $V_C(t)$ and $V_D(t)$ are applied to a
   gate in case b) to move a localized level with energy $E(t)$ up
   (down) through the Fermi energy releasing an electron (hole), or in
   case c) to cycle the transmission probability, $T(t)$, from zero
   through unity and back to generate a particle-hole excitation.}
\end{figure}

Here we propose an entanglement pump, aiming for the simplest
scheme for
detection and optimal production of orbital entanglement
in the same system, see
Fig.\/ \ref{fig:fig1}. A conductor in the quantum Hall
regime with two on-demand sources, C and D, is connected to four 
terminals via electronic beam-splitters at A and B. Using 
a single spin-polarized edge state and the interferometer of
\cite{Samuelsson05}, the sources operate in the
strong amplitude regime and generate pairs of
entangled particles at an optimal rate.

The two types of sources, driven by voltage pulses $V_C(t)$ and
$V_D(t)$, are shown in Figs. \ref{fig:fig1}(b) and (c). 
Fig. \ref{fig:fig1}(b) shows the level of a quantum dot (QD) driven up
(down) through Fermi energy, generating a single electron (hole),
as in the  experiment of \cite{Feve}. However, with the 
QD coupled to separate quantum Hall edge states, the particle is
emitted into a superposition of states at the two edges. In
Fig. \ref{fig:fig1}(c), a quantum point contact (QPC) is opened and
closed to generate exactly one particle and one hole, which are independently
emitted into edge state superpositions. We show below that, for
QD as well as QPC sources, driving C and D periodically generates
orbitally entangled pairs of both electron and hole wavepackets, with
one particle propagating towards A and one towards B. For synchronized
and spatially symmetric sources the entanglement production rate is
optimal, with half a Bell pair per cycle. Using earlier results for
many-body states of emitted particles of on-demand sources
\cite{Keeling06,Keeling08,Sherkunov09}, we derive explicit expressions
for the entangled wavefunction. The entanglement, arising from
two-particle interference \cite{Sam04,Ned07,Sam09}, can be detected
via low-frequency cross correlations of currents at the four terminals
1-4. Throughout the paper consider zero temperature and put
$\hbar,e=1$.

{\it QD-sources:} We first consider QD-sources. 
The aim is to obtain the
many-body wavefunction of the entangled particles emitted from the
sources towards A and B. We  consider the scattering
properties of source C for a single electron emission event. The QD has a
single localized level at energy $E(t)$ and tunnel couplings
to the
edges towards A and B. 
If the energy of the level is varied slowly on the scale of the Wigner
delay time (adiabatic approximation), the time-development of the 
scattering states is given by the instantaneous value of the
scattering matrix, $S_C$. When the energy of the level is
varied at constant speed, $E(t)=\nu (t-t_C)$, $S_C$ 
takes the Breit-Wigner form:
\begin{eqnarray} 
  S_C = \frac{1}{t-t_C-i\tau_{C}}
  \left( \begin{array}{cc}
      t-t_C+i\bar \tau_{C} &-2i\sqrt{\tau_{CA} \tau_{CB}} \\
      -2i\sqrt{\tau_{CA} \tau_{CB}} & t-t_C-i\bar \tau_{C}\end{array} \right)
\label{Sdot}
\end{eqnarray} 
with $\tau_{C}=\tau_{CA}+\tau_{CB}$ and $\bar
\tau_C=\tau_{CA}-\tau_{CB}$. Here $\nu \tau_{CA}$
and $\nu\tau_{CB}$ are tunnel rates
through the barriers into states propagating towards A and B. 

The scattering
matrix, $S_C$, can be diagonalized in a time-independent basis,
$\tilde{S}_C=VS_CV^{\dagger}=\left( \begin{array}{ccc}
e^{i\phi_C} & 0 \\
0 & 1\end{array} \right)$, with $V = \frac{1}{\sqrt{\tau_{C}}} \left( \begin{array}{cc}
-\sqrt{\tau_{CA}}  & \sqrt{\tau_{CB}} \\
\sqrt{\tau_{CB}} & \sqrt{\tau_{CA}} \end{array} \right)$ and 
\begin{equation}
e^{i\phi_C(t)}=\frac{t-t_C+i\tau_C}{t-t_C-i\tau_C}.
\label{phase}
\end{equation}
The low temperature many-body state incident on C is a filled Fermi
sea of electrons originating from terminals 2 and 3. 
The incident many-body state is $|\psi_{in}\rangle
=\prod_{\epsilon <0} a_{C+}^{\dagger}(\epsilon )
a_{C-}^{\dagger}(\epsilon)|\rangle$, where the
$a_{C\pm}^{\dagger}(\epsilon)$ are creation operators in the diagonal
basis for particles with energy $\epsilon$  in the incoming channels $+$ or $-$,
and $ |\rangle$ is the vacuum.
Denoting the operators for
particles in the  corresponding outgoing states by $b_{C\pm}^{\dagger}(\epsilon)$, 
the many-body state after impinging on C is
\begin{equation} 
|\psi_C\rangle =\prod_{\epsilon
  <0} b_{C+}^\dagger(\epsilon)  \sum_{\epsilon '}
(e^{i\phi_C})_{\epsilon , \epsilon'} 
b^\dagger_{C-}(\epsilon')|\rangle.
\label{psiout}
\end{equation}
Here $(e^{i\phi_C})_{\epsilon,\epsilon'}$ is
the Fourier transform of
$e^{i\phi_C(t)}$ with respect to the energy difference
$(\epsilon-\epsilon')$ \cite{AdamovMuzykantskii,buttpump,dAMuz05}.

With the phase profile of (\ref{phase}) and taking $\nu>0$,
the many-body state  becomes $|\psi_C^e\rangle
=\sqrt{2\tau_C}\sum_{\epsilon>0}\exp[\epsilon(it_C-\tau_C)]b^{\dagger}_{C-}(\epsilon)|0\rangle$. It
describes a single electron 
above the unperturbed Fermi sea $|0\rangle$
\cite{Keeling06,Keeling08}.
The
wavefunction of the single particle excitation is $\psi(x,t) \sim
1/(x-v_{dr}[t-t_C-i\tau_C])$, where $x$ is the distance from C and
$v_{dr}$ the drift velocity along the edge. 
Using
$(b_{CA}^{\dagger}(\epsilon),b_{CB}^{\dagger}(\epsilon))=V(b_{C-}^{\dagger}(\epsilon),b_{C+}^{\dagger}(\epsilon))$
to write $|\psi_C^e\rangle$ in
terms of operators $b_{CA}^{\dagger}(\epsilon)$ and
$b_{CB}^{\dagger}(\epsilon)$, which create particles propagating along the
edges towards A and B, we obtain
\begin{eqnarray}
  |\psi^e_C\rangle=\frac{1}{\sqrt{\tau_C}}\left[-\sqrt{\tau_{CA}}p^{\dagger}_{CA}+\sqrt{\tau_{CB}}p^{\dagger}_{CB}\right]|0\rangle. 
\label{psiC}
\end{eqnarray}
Here the normalized wavepacket operator
$p^{\dagger}_{CA}=\sqrt{2\tau_C}\sum_{\epsilon>0}\exp[\epsilon(it_C-\tau_C)]b^{\dagger}_{CA}(\epsilon)$
and similarly for $p_{CB}^{\dagger}$. Eq (\ref{psiC}) shows
that driving the QD-level at C up through the Fermi
level injects a single electron into a linear superposition of
states propagating towards A and B.

The corresponding result to  (\ref{psiC}), when driving the QD
level at D
up through the Fermi level, is
$|\psi^e_D\rangle=(1/\sqrt{\tau_D})\left[-\sqrt{\tau_{DA}}e^{-i\eta}p^{\dagger}_{DA}+\sqrt{\tau_{DB}}e^{i\eta}p^{\dagger}_{DB}\right]|0\rangle$. Here
$p^{\dagger}_{DA}$ is obtained from $p^{\dagger}_{CA},p^{\dagger}_{DB}$ by changing
indices $C\leftrightarrow D$.  We have
also included a phase-factor $e^{i\eta}$ to account for a possible
Aharonov Bohm phase. The total wave function for the electrons emitted
from C and D is then the product
$|\psi_{out}^{ee}\rangle=|\psi^e_C\rangle |\psi^e_D\rangle$. We are
interested in the non-local properties of $|\psi_{out}^{ee}\rangle$,
with the particles emitted towards different beamsplitters. 
Projecting $|\psi_{out}^{ee}\rangle$ onto the subspace
with one particle at A and one at B, gives the normalized
wavefunction \cite{Beenrev}
 \begin{eqnarray}
 |\psi^{ee}_{AB}\rangle &=&\frac{1}{\sqrt{N}}
 \left[\sqrt{\tau_{CA}\tau_{DB}}p^{\dagger}_{CA}p^{\dagger}_{DB}e^{i\eta} \right. \nonumber \\
  &+& \left.\sqrt{\tau_{CB}\tau_{DA}}p^{\dagger}_{CB}p^{\dagger}_{DA}e^{-i\eta}\right]|0\rangle
 \end{eqnarray}
 where $N=\tau_{CA}\tau_{DB}+ \tau_{CB}\tau_{DA}$. The weight, or
 probability that $ |\psi^{ee}_{AB}\rangle$ is generated, is
 $w_{AB}=N/(\tau_C\tau_D)$. 

 The wavefunction $|\psi^{ee}_{AB}\rangle$ describes two particles
 above the filled Fermi sea. It is entangled in the orbital
({\it  i.e.\/} source C and D) degree of freedom---a result of two-particle
 interference \cite{Sam04}. The superposition in
 $|\psi^{ee}_{AB}\rangle$ results from the indistinguishability of the
 two emission processes leading to a particle at A and one at B. In
 one process a particle moves from C to A and another from D to B.
 In the second process particles move from D to A and from C to B. To
 quantify the orbital entanglement we consider the reduced $4\times 4$
 orbital density matrix $\rho_{AB}$. This is obtained by tracing
 $|\psi^{ee}_{AB}\rangle \langle\psi^{ee}_{AB}|$ over energy \cite{Samrenorm}:
\begin{eqnarray}
 \rho_{AB}&=&N^{-1}\big( \tau_{CA}\tau_{DB}|CD\rangle\langle
   CD|+\tau_{CB}\tau_{DA}|DC\rangle\langle DC| \nonumber \\
&+&  \chi\sqrt{\tau_{CA}\tau_{CB}\tau_{DA}\tau_{DB}}
  \left[e^{i\eta}|CD\rangle\langle DC|+h.c.\right] \big)
\label{reddens}
\end{eqnarray}
where $|CD\rangle \equiv |C\rangle_A|D\rangle_B$ and $|C\rangle_A$
denotes an electron at A emitted from C etc. Here 
\begin{eqnarray}
\chi = \frac{4\tau_{C}\tau_{D}}{(t_C-t_D)^2+(\tau_C+\tau_D)^2}
\label{chiFunction}
\end{eqnarray}
quantifies the overlap ($0 \leq \chi \leq 1$)  between the two wavepackets emitted from C
and D.
We note that a reduced overlap $\chi<1$ plays the same role as a non-zero
dephasing for two-particle interference \cite{Sam09}. 

The entanglement of $\rho_{AB}$ is conveniently quantified via the
concurrence ${\cal C}$, which ranges from $0$, for a
separable non-entangled state, to $1$ for a maximally entangled
state \cite{Wooters} . For $\rho_{AB}$ we find
\begin{equation}
{\cal C}=2(\chi/N)\sqrt{\tau_{CA}\tau_{CB}\tau_{DA}\tau_{DB}}
\label{conc}
\end{equation}
which is non-zero for arbitrary small overlap $\chi$ and
arbitrary tunnel couplings \cite{Olkhovskaya}. For QDs
synchronized in time, $t_C=t_D$, with symmetric couplings
$\tau_{CA}=\tau_{CB}=\tau_{C}/2$ and $\tau_{DA}=\tau_{DB}=\tau_{D}/2$,
the density matrix
$\rho_{AB}=|\varphi^{ee}_{AB}\rangle\langle\varphi^{ee}_{AB}|$ with
$|\varphi^{ee}_{AB}\rangle=1/\sqrt{2}[e^{i\eta}|CD\rangle+e^{-i\eta}|DC\rangle]$.
This
state is an orbital Bell pair, {\it i.e.\/} it is maximally entangled  (${\cal
  C}=1$). Moreover, the weight $w_{AB}=1/2$ and hence the concurrence
production per cycle is $w_{AB}{\cal C}=1/2$, the
theoretical maximum  \cite{Beenrev}.

If the QD-level at C is driven down through the Fermi level
at time $t_C'$, a hole is generated in a linear superposition of
states in the edges towards A and B, with a wavefunction
$|\psi^h_C\rangle=(1/\sqrt{\tau_C})\left[-\sqrt{\tau_{CA}}h^{\dagger}_{CA}+\sqrt{\tau_{CB}}h^{\dagger}_{CB}\right]|0\rangle$. Here
the hole wavepacket creation operator is
$h^{\dagger}_{CA}=\sqrt{2\tau_C}\sum_{\epsilon<0}\exp[\epsilon(it_C'+\tau_C)]b_{CA}(\epsilon)$. A
similar relation holds for $|\psi^h_D\rangle$, giving a total hole wavefunction
$|\psi^{hh}_{out}\rangle=|\psi^h_C\rangle|\psi^h_D\rangle$.

For a large amplitude driving up and subsequently down through the
Fermi energy, with $t_C'-t_C \gg \tau_C$ and $t_D'-t_D \gg \tau_D$,
the electron and hole emissions are well separated in time and can be
treated as independent \cite{Mosk08,Keeling08}. The total wavefunction
for the emitted particles is then
$|\psi^{QD}_{out}\rangle=|\psi^{ee}_{out}\rangle|\psi^{hh}_{out}\rangle$. In
the experimentally relevant situation \cite{Feve}, with a cycling of
the QD-level with period $T \gg \tau_C,\tau_D$, the pump produces
pairs of both entangled electrons and holes which, for symmetric and
syncronized sources, reaches the optimal production rate of $1/2$ a Bell
pair per period.

{\it QPC-sources.} For QPC sources, the emitted state is created by
applying tailored voltage pulses $V_{C,D}(t)$ to the gates C and D
\cite{Sherkunov09}. The instantaneous scattering matrix of gate C can be
written
\begin{equation} 
  S_C=  \left( \begin{array}{ccc}
      \lambda_C(t) & \kappa_C(t) \\
      - \kappa_C^{*}(t) & \lambda_C^{*}(t) \end{array} \right)  
\label{Stunnel}
\end{equation} 
where $|\kappa_C(t)|^2$ and 
$|\lambda_C(t)|^2$
are the transmission and reflection  probabilities through the
QPC respectively. Provided no voltage is applied across the device, 
$S_C$ can be diagonalized \cite{Sherkunov09}
in a time-independent basis $\tilde S_C=V S_C V^{\dagger}=\left( \begin{array}{ccc}
e^{i\phi_{C}}& 0 \\
0 & e^{-i\phi_{C}} \end{array} \right) $,  with
$V = \frac{1}{\sqrt{2}} \left( \begin{array}{cc}
1  & -i \\
1 &  i  \end{array} \right)$
and $e^{i\phi_{C}(t)} = \lambda_C(t) + i\kappa_C(t)$. The outgoing
manybody wavefunction can then be written in the diagonal basis
\begin{eqnarray}
 |\bar \psi_C\rangle = \frac{1}{2}\prod_{\epsilon <0}\sum_{\epsilon ',
   \epsilon ''}(e^{i\phi_C })_{\epsilon , \epsilon '}(e^{-i\phi_C})_{\epsilon , \epsilon ''} b^{\dagger}_{-}(\epsilon')b^{\dagger}_{+}(\epsilon '')
 |\rangle. 
\label{psioutCTunnel}
 \end{eqnarray}
 By varying $V_{C}(t)$ so that the transmission amplitude
 $\kappa_C(t)=2\tau_C(t-t_C)/[(t-t_C)^2+\tau_C^2]$, the phase factor
 $e^{i\phi_C(t)}$, is again given by (\ref{phase}). In this case we can
write the outgoing manybody wavefunction  \cite{Keeling06} 
\begin{eqnarray}
 |\psi_C^{eh}\rangle=\frac{1}{2}\left(p^{\dagger}_{CA}+ip^{\dagger}_{CB}\right)\left(h^{\dagger}_{CA}-ih^{\dagger}_{CB}\right)|0\rangle
\end{eqnarray}
in the unrotated basis with a corresponding result for
$|\psi_D^{eh}\rangle$. 
The wavefunctions $ |\psi_{C,D}^{eh}\rangle$
describes one electron and one hole wavepacket, independently emitted
on top of the unperturbed Fermi sea, in superpositions which describe
excitations propagating
towards A and B. Consequently, the total state
$|\psi^{QPC}_{out}\rangle=|\psi^{eh}_{C}\rangle|\psi^{eh}_{D}\rangle$,
as for the QD, is the direct product of independently emitted
pairs of electrons and holes. The difference is that for the QPC the
electrons and holes at C (D) are emitted at the same time, $t_C'=t_C$
 and $t_D'=t_D$. The reduced density matrix and the concurrence of the
electron and hole states at A and B are given by Eqs.\/ (\ref{reddens})
and (\ref{conc}) respectively, after taking
$\tau_{CA}=\tau_{CB}=\tau_{C}/2$ and
$\tau_{DA}=\tau_{DB}=\tau_{D}/2$. For perfectly synchronized driving
$t_C=t_D$ the QPC  also works as an optimal entanglement pump,
producing independent pairs of entangled electrons and holes at A and
B with a rate $1/2$ per cycle.

{\it Bell inequality} The existence of entanglement in the system
illustrated in Fig. \ref{fig:fig1} can be verified experimentally by demonstrating
violation of a Bell
inequality. A BI can conveniently be formulated in terms of the joint,
or coincident, probabilities to detect quasiparticle excitations
\cite{Sam04}, as $M \leq 2$ for the Bell parameter
\begin{equation}
M = |E(\alpha,\beta)-E(\alpha,\beta ')+E(\alpha ',\beta)+E(\alpha
',\beta ')|.
\label{M}
\end{equation}
Here the correlation function
\begin{eqnarray}
E(\alpha,\beta) & = & \frac{P_{24}+P_{13}-P_{14}-P_{23}}{P_{24}+P_{13}+P_{14}+P_{23}},
\label{E}
\end{eqnarray}
where $P_{ij}$ ($=P_{ij}^{ee}+P_{ij}^{hh}$) is the sum of the probabilities
to jointly detect an electron/hole at A ($i=1,2$) and B ($i=3,4$)
during a pumping cycle \cite{Samuelsson05}, and is given by
\begin{eqnarray}
 P_{ij}^{ee}=\int_0^{\infty}d\epsilon \int_0^{\infty} d\epsilon '\langle b_i^{\dagger}(\epsilon)b_j^{\dagger}(\epsilon ')b_j(\epsilon ') b_i (\epsilon)\rangle,\nonumber\\
 P_{ij}^{hh}=\int_{-\infty}^0 d\epsilon \int_{-\infty}^0 d\epsilon'\langle b_i(\epsilon)b_j(\epsilon ')b_j^{\dagger}(\epsilon ') b_i^{\dagger}(\epsilon)\rangle.
\label{p}
\end{eqnarray}
The annihilation operators for the states entering the terminals $i,j$
at A are 
\begin{eqnarray}
 \left(\begin{array}{ll} b_{2} \\  b_{1}\end{array}\right)= \left( \begin{array}{ccc}
\cos\alpha & \sin\alpha \\
-\sin\alpha & \cos\alpha \end{array} \right)  \left(\begin{array}{ll}
b_{CA} \\  b_{DA}\end{array}\right) .
\label{b1} 
\end{eqnarray}
The operators at $B$ are obtained after setting $\alpha \rightarrow \beta$, $1\rightarrow 3$ and $2\rightarrow 4$. From the
emitted manybody wavefuctions for the QD and the QPC, together with the
operator relations at the beamsplitters (\ref{b1}), we obtain
the joint detection probability 
\begin{eqnarray}
&&P^{ee}_{13}=(2\tau_{CA}\tau_{DB}\sin^2\alpha \cos^2\beta+ 2\tau_{CB}\tau_{DA}\sin^2 \beta \cos^2 \alpha \nonumber\\
&&-\chi \cos \eta \sqrt{\tau_{CA}\tau_{DB}\tau_{CB}\tau_{DA}} \sin 2\alpha \sin 2 \beta)/(2\tau_C\tau_D)
\label{Pform} 
\end{eqnarray}
with corresponding results for the other $P^{ee}_{ij}$ ($P_{ij}^{hh}=P_{ij}^{ee}$)
\cite{Floquet}. For the QPC-source we put
$\tau_{CA}=\tau_{CB}=\tau_C/2$ and
$\tau_{DA}=\tau_{DB}=\tau_D/2$. Inserting the $P_{ij}$ into
(\ref{E}) gives
\begin{equation}
E(\alpha,\beta)   = -(  \cos2\alpha\cos2\beta +
  {\cal C}\cos \eta \sin2\alpha\sin2\beta) 
\label{Eform}
\end{equation}
where ${\cal C}$ is the concurrence in (\ref{conc}). By optimising
the settings $\alpha$, $\alpha'$, $\beta$ and $\beta'$ of the
beamsplitters A and B \cite{Samuelsson03}, the BI reduces to
$2\sqrt{1+{\cal C}^2\cos^2 \eta}\le 2$ which can be violated for arbitrary
overlap $\chi$ and phase $\eta$. 

The joint detection probabilities in (\ref{p}) are presently not
directly accessible in mesoscopic conductors, as they require 
time-resolved correlation measurements on the time scale of the period
$T$. It is however possible to express $P_{ij}$ in terms of
experimentally available currents and low frequency current
cross-correlators. The period-averaged, zero frequency
cross-correlations $ S_{ij}$ \cite{BlanterButtiker} are found to be
\begin{eqnarray}
  S_{ij}= T^{-1}\left(P_{ij}^{ee}-P^e_iP^e_j+P_{ij}^{hh}-P^h_iP^h_j \right).
\label{CurrentCorrelation}
\end{eqnarray}
Note that $ S_{ij}$ does not contain any electron-hole correlations,
as a
consequence of the independent emission of electrons and holes,
discussed above. In (\ref{CurrentCorrelation})
the (single particle) probability to detect an
electron in lead $i$ is $P^e_i=\int_0^{\infty}d\epsilon\langle b_i^{\dagger}(\epsilon)
b_i(\epsilon)\rangle$, while $P^h_i$ is the probability to detect a hole.
For the QD $P^e_i$ and $P^h_i$ are available from  $I_i(t)$,
which is the
experimentally accessible time-resolved current \cite{Feve}.
By integrating over half-cycles, driving the QD-level up (down) during
the first (second) half, we have $(1/e)\int_0^{T/2} dt
I_i(t)=Q_i^e/e=P_i^e$ and $(1/e)\int_{T/2}^{T} dt
I_i(t)=Q_i^h/e=P_i^h$. Here $Q^e_i$ and $Q^h_i$ are the electron and
hole charge flowing into contact $i$ during the cycle. For the QPC
$P^e_i$ and $P^h_i$ can instead be obtained in a more indirect way via
the low-frequency current autocorrelations, which we do
not discuss here. Taken together, this allows us to express the correlation
function $E(\alpha,\beta)$, and hence the Bell inequality, directly in
terms of currents and current correlations.

In conclusion, we have proposed an optimal entanglement on-demand
source in a non-interacting mesoscopic conductor in the quantum Hall
regime. Pairs of entangled, spatially separated electrons and holes
are generated by applying tailored, time-dependent voltage pulses to
quantum dots or quantum point contacts. The entanglement can be
investgated by current and current cross correlation measurements.

We acknowledge discussions with B. Muzykantskii.

\bibliographystyle{apsrev}
\bibliography{entanglerRESUBnda}

\begin{thebibliography}{25}
\bibitem[{Flo()}]{Floquet}
\bibinfo{note}{We point out that the same result is obtained by employing a
  Floquet scattering approach \cite{buttpump}, where no use is made of the
  wavefunctions $|\Psi_{out}^{QD}\rangle$ or $|\Psi_{out}^{QPC}\rangle$.}
\end{thebibliography}

\end{document}